\documentclass{aa}

\usepackage{txfonts}
\usepackage{graphicx}

\begin{document}

\title{Visual orbit for the low-mass binary 
       Gliese~22~AC from speckle interferometry 
       \thanks{Based on observations collected at the German-Spanish
        Astronomical Centre on Calar Alto, Spain}}

\author{Jens Woitas \inst{1, 3} \and
        Vakhtang S. Tamazian \inst{2} \and
        Jos\'e A. Docobo \inst{2} \and
        Christoph Leinert \inst{3}}

\institute{Th\"uringer Landessternwarte Tautenburg, Sternwarte 5,
           D-07778 Tautenburg, Germany \and
           Observatorio Astron\'omico Ram\'on Mar\'{\i}a Aller,
           Universidade de Santiago de Compostela, Avenida das Ciencias s/n,
           15782 Santiago de Compostela, Spain \and
           Max-Planck-Institut f\"ur Astronomie, K\"onigstuhl 17,
           D-69117 Heidelberg, Germany}

\offprints{Jens Woitas \email{woitas@tls-tautenburg.de}}

\date{Received / Accepted}

\abstract{
Based on 14 data points obtained with near-infrared speckle interferometry
and covering an almost entire revolution, 
we present a first visual orbit for the low-mass
binary system Gliese~22~AC. The quality of the orbit is largely improved
with respect to previous astrometric solutions. The dynamical system
mass is $0.592\pm 0.065\,M_{\sun}$, where the largest part of the
error is due to the {\it Hipparcos} parallax. A comparison of this
dynamical mass with mass-luminosity relations on the lower main sequence
and theoretical evolutionary models for low-mass objects shows
that both probably underestimate the masses of M~dwarfs. A mass
estimate for the companion Gliese~22~C indicates that this object is a
very low-mass star with a mass close to the hydrogen burning mass limit. 

          }

\titlerunning{Visual orbit for Gliese~22~AC}
\authorrunning{Woitas et al.}
\maketitle

\section{Introduction}
M~dwarfs are the dominant population of the Galaxy as well in
numbers as in stellar mass contribution. Furthermore, they mark
the transition regime between stars and the now well established
classes of substellar objects. Given these important properties, it
is problematic that there is only a small number of empirically
determined stellar masses at the lower end of the main sequence.
In addition, most of these dynamical masses are affected with
large uncertainties (e.\,g. Henry \cite{Hen98}). This means that
the mass-luminosity relation for M~dwarfs is not well calibrated.
Moreover, theoretical evolutionary models for low-mass objects
that cover also the substellar regime are not well checked
with dynamical masses.\\
As a contribution to a solution of these problems we are carrying
out a program aiming at a determination of visual orbits and thus
dynamical masses for M~dwarf binaries. Speckle interferometry
with array cameras in the near infrared allows highly precise
measurements of the relative astrometry in sub-arcsecond binary
or triple systems.\\
This program has already led to orbit determinations for the very
low-mass systems \object{Gliese~866} (Woitas et al.\,\cite{Woi00}) and 
\object{LHS~1070}~BC (Leinert et al.\,\cite{Lei01}). The companions
in these systems appear to have masses close to the stellar/substellar
limit at $M\approx 0.075\,M_{\odot}$.
In this paper we discuss a first visual orbit determination
for \object{Gliese~22}~AC that was briefly announced in
IAU Commission 26 Information Circular 147 (Docobo et al.\,\cite{Doc02}).\\
A visual companion to the M2 star Gliese 22 (other designations:
\object{HIP 2552}, \object{BD +66$^{\circ}$34}, \object{ADS 440},
\object{V~547~Cas}) was first reported by Espin \& Milburn (\cite{Esp26}).
At the time of this detection the projected separation was 2\farcs79.
The orbital motion of this companion with a period of $\approx 320$~yr
has been monitored since its detection, and most recent orbital elements
are given by Lampens \& Strigachev (\cite{Lam01}). Alden (\cite{Ald47})
found that the primary component of this pair is itself an astrometric
binary. Herafter, we will refer to this close pair as Gliese~22~AC and
to the more distant third component as Gliese~22~B. Hershey (\cite{Her73})
presented orbital elements for Gliese~22~AC based on a rich
collection of astrometric plates from the Sproul Observatory.
This calculation was refined by Heintz (\cite{Hei93}) adding
more data points and S\"oderhjelm (\cite{Sod99}) including the
{\it Hipparcos} parallax of Gliese~22 into the analysis.The first two resolved
{\it visual} observations of Gliese~22~AC were reported by McCarthy
et al. (\cite{Mcc91}) using near-infrared speckle interferometry.
Since the epoch of these first two measurements we regularly observed
this pair obtaining 12 more data points, which uniformly cover almost
an entire revolution. Based on these data, we present in this paper
a visual orbit and a dynamical system mass that are 
more precise than the previous astrometric solutions.
We will describe the techniques of observations and data analysis
in Sect.\,\ref{obs} and present the result of the orbit calculation
in Sect.\,\ref{results}. In Sect.\,\ref{disc} we will discuss implications
of the derived dynamical system mass on the mass-luminosity relation
and theoretical models for very low-mass objects.

\section{Observations and Data Analysis}
\label{obs}

\begin{figure*}
\sidecaption
\includegraphics[width=6cm]{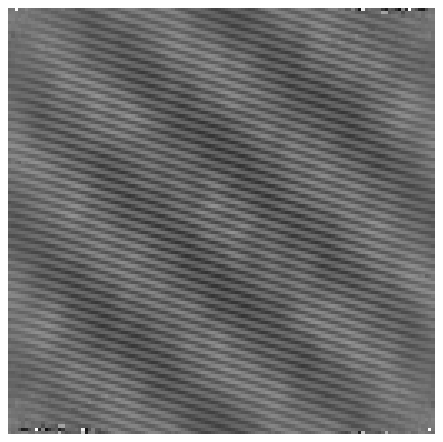}
\includegraphics[width=6cm]{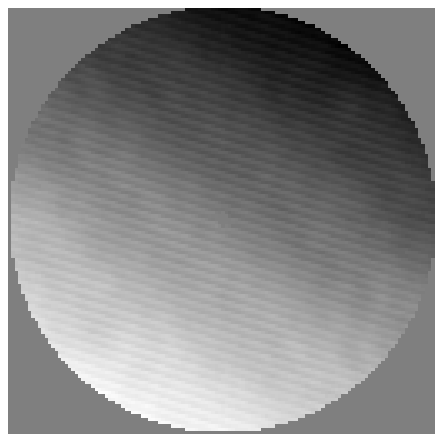}
\caption{\label{vispha} Modulus (left) and bispectrum-phase (right) for
 Gliese~22, derived from 1000 short ($\tau = 0.2\,\mathrm{s}$) exposures
 obtained at $\lambda = 2.2\,\mu\mathrm{m}$ with OMEGA Cass at the 3.5-m
 telescope on Calar Alto on 3 November 2001. North is to the top and east
 to the left. The broad stripe pattern in the modulus is due to Gliese~22~AC
 while the fine patterns (roughly from top to bottom) are produced by
 the more distant companion Gliese~22~B. The phase has been cut off
 at the Nyquist frequency that is 5.3~arcsec$^{-1}$ for the adopted
 pixel scale of 0\farcs095/pixel.}
\end{figure*}

\begin{table*}
\centering
\caption{\label{obs_AC_table} Overview of all observations of Gliese~22~AC
 used for the visual orbit determination in this paper. Except for the
 first two data points all observations have been carried out at the
 3.5-m telescope on Calar Alto.}
\begin{tabular}{lllllllll}
No. & Date & Epoch & Position angle & Projected & Weight & Filter &
 Flux ratio & Instrument \\ 
    &      &       &  angle [$^{\circ}$] & separation [mas] & & &
 $F_C/F_A$ & (or reference) \\ \hline
1 & 12.10.1989 & 1989.7803 & 42.7 $\pm$ 2.5 & 451 $\pm$ 20 & 4 & K & 
  0.167 $\pm$ 0.009 & McCarthy et al.\,\cite{Mcc91} \\
2 & 10.12.1989 & 1989.9418 & 43.9 $\pm$ 2.5 & 453 $\pm$ 20 & 4 & H & 
  0.143 $\pm$ 0.008 & McCarthy et al.\,\cite{Mcc91} \\
3 & 11.09.1990 & 1990.6954 & 57.1 $\pm$ 6.3 & 405 $\pm$ 39 & 5 & K & 
  & 1D  \\
4 & 20.09.1991 & 1991.7201 & 71.6 $\pm$ 1.6 & 443 $\pm$ 13 & 5 & K & 
  & 1D  \\
5 & 30.09.1993 & 1993.7474 & 122.5 $\pm$ 0.5 & 464 $\pm$ 11 & 7 & K &
  0.151 $\pm$ 0.004 & MAGIC \\
6 & 28.01.1994 & 1994.0767 & 129.3 $\pm$ 4.3 & 478 $\pm$ 40 & 10 & K &
  & MAGIC \\
7 & 15.09.1994 & 1994.7064 & 141.1 $\pm$ 0.4 & 510 $\pm$ 5 & 9 & K &
  0.141 $\pm$ 0.007 & MAGIC \\
8 & 13.12.1994 & 1994.9501 & 145.2 $\pm$ 0.2 & 532 $\pm$ 5 & 9 & K &
  & MAGIC \\
9 & 08.10.1995 & 1995.7693 & 157.5 $\pm$ 1.7 & 526 $\pm$ 16 & 9 & K &
  0.158 $\pm$ 0.004 & MAGIC \\
10 & 28.09.1996 & 1996.7448 & 172.1 $\pm$ 0.2 & 533 $\pm$ 4 & 10 & K &
  0.141 $\pm$ 0.004 & MAGIC \\
11 & 17.11.1997 & 1997.8789 & 190.2 $\pm$ 0.3 & 460 $\pm$ 4 & 7 & K &
  0.171 $\pm$ 0.008 & MAGIC \\
12 & 09.02.2001 & 2001.1095 & 302.3 $\pm$ 0.3 & 334 $\pm$ 4 & 9 & K &
  0.185 $\pm$ 0.005 & OMEGA Cass \\
13 & 03.11.2001 & 2001.8406 & 325.5 $\pm$ 0.5 & 402 $\pm$ 6 & 8 & K &
  0.151 $\pm$ 0.011 & OMEGA Cass \\
14 & 19.10.2002 & 2002.7994 & 348.2 $\pm$ 0.4 & 463 $\pm$ 5 & 9 & K &
  0.150 $\pm$ 0.005 & OMEGA Cass \\
\end{tabular}
\end{table*}

\begin{table*}
\centering
\caption{\label{obs_AB_table} New observations of the wide pair 
 Gliese~22~AB obtained at the 3.5-m telescope on Calar Alto.}
\begin{tabular}{llllllll}
No. & Date & Epoch & Position angle & Projected & Filter & Flux ratio 
 & Instrument \\ 
    &      &       &  angle [$^{\circ}$] & separation & & $F_B/F_A$ &  \\ \hline
1   & 30.09.1993 & 1993.7474 &  164.4 $\pm$ 0.5 & $4\farcs24 \pm 0\farcs09$ &
 K & 0.199 $\pm$ 0.004 & MAGIC \\
2   & 28.01.1994 & 1994.0767 &  165.3 $\pm$ 0.6 & $4\farcs28 \pm 0\farcs06$ &
 K & & MAGIC \\
3   & 15.09.1994 & 1994.7064 &  167.1 $\pm$ 0.4 & $4\farcs29 \pm 0\farcs04$ &
 K & & MAGIC \\  
4   & 13.12.1994 & 1994.9501 &  166.9 $\pm$ 0.1 & $4\farcs29 \pm 0\farcs01$ &
 K & 0.267 $\pm$ 0.004 & MAGIC \\
5   & 08.10.1995 & 1995.7693 &  167.4 $\pm$ 0.2 & $4\farcs29 \pm 0\farcs01$ &
 K & & MAGIC \\  
6   & 28.09.1996 & 1996.7448 &  168.7 $\pm$ 0.1 & $4\farcs28 \pm 0\farcs01$ &
 K & 0.245 $\pm$ 0.004 & MAGIC \\
7   & 17.11.1997 & 1997.8789 &  170.0 $\pm$ 0.3 & $4\farcs23 \pm 0\farcs01$ &
 K & 0.325 $\pm$ 0.004 & MAGIC \\  
8   & 09.02.2001 & 2001.8406 &  172.8 $\pm$ 0.2 & $4\farcs00 \pm 0\farcs01$ &
 K & 0.246 $\pm$ 0.006 & OMEGA Cass \\ 
9   & 03.11.2001 & 2001.8406 &  172.8 $\pm$ 0.1 & $3\farcs99 \pm 0\farcs01$ &
 K & 0.270 $\pm$ 0.009 & OMEGA Cass \\
10  & 19.10.2002 & 2002.7994 &  173.0 $\pm$ 0.1 & $3\farcs956 \pm 0\farcs005$
 & K & 0.298 $\pm$ 0.005 & OMEGA Cass \\ 
\end{tabular}
\end{table*}

The database for our visual orbit determination for Gliese~22~AC
is given in Table\,\ref{obs_AC_table}. The observations numbered 1 and 2
were taken from McCarthy et al. (\cite{Mcc91}) while the other measurements
are published here for the first time. Observations 3 and 4 made use
of one-dimensional speckle interferometry. This observing technique and
the reduction of these data are described in Leinert \& Haas
(\cite{Lei89}).\\
All other data points have been obtained with the near-infrared
cameras MAGIC and OMEGA Cass at the 3.5-m telescope on Calar Alto.
Both instruments are capable of taking fast sequences of short time
exposures ($t_{\mathrm{exp}}\approx 0.1\,\mathrm{s}$) and in this
way allow speckle interferometry with two-dimensional detector arrays.
Typically we have taken 1000 short exposures for Gliese~22
and the nearby PSF calibrator (single star) \object{SAO~11358}
in the K band ($\lambda = 2.2\,\mu\mathrm{m}$).
A detailed overview of the data reduction and analysis has been given
by K\"ohler et al. (\cite{Koe00}). Briefly, we obtain the modulus of
the complex visibility by deconvolving the power spectrum of
Gliese~22 with that of the PSF calibrator. The phase is
recursively reconstructed using the algorithm by Knox \& Thompson 
(\cite{KT74}) and the bispectrum method (Lohmann et al. \cite{Loh83}).
As an example we show in Fig.\,\ref{vispha} modulus and bispectrum phase
obtained from the observation at 3 Nov 2001.
In the two-dimensional observations Gliese~22~B is also in the
detector array. Therefore, we fit a triple star model to the complex
visibility and in this way determine the relative astrometry and the
flux ratios of the companions B and C with respect to Gliese~22~A.
Pixel scale and detector orientation are derived from astrometric fits
to images of the Orion Trapezium cluster core where precise astrometry has
been given by McCaughrean \& Stauffer (\cite{Mcc94}). These calibration
observations are however only available for the observations since
1995 (No. 9 to 14 in Table \ref{obs_AC_table}). For the previous
observations we have determined pixel scale and detector orientation
from visual binary stars that either have well known orbits
(\object{$\alpha$ Psc}, \object{$\zeta$ Aqr}) or show no measurable
orbital motion within 10\,yr (\object{RNO~1~BC}). Position angles
and projected separations of these binaries were calibrated with
the help of the Trapezium cluster in subsequent observing runs.
As an example we show our Trapezium-calibrated measurements
of $\alpha$ Psc in Table~\ref{ast-calib}, together with the
prediction of the ephemerides from Scardia (\cite{Sca83}).
In this way all our two-dimensional speckle observations were placed
into a consistent system of pixel scale and detector orientation.\\
Position angles, projected separations and flux ratios for the
companions C and B are given in Tables \ref{obs_AC_table} and
\ref{obs_AB_table}. The formal uncertainties of the relative astrometry
are typically 5 -- 10 milli-arcsec in $x$ and $y$ for the 2D data.
These errors seem to be reasonable since they are in the same order of
magnitude as the residuals of the orbital fit (see Table\,\ref{results}).

\begin{table*}
\centering
\caption{\label{ast-calib} Our observations of $\alpha$ Psc that employ
 an astrometric calibration derived from images of the Trapezium
 cluster core, compared to the prediction of the ephemerides from
 Scardia (\cite{Sca83}). The mean residuals were used as a
 correction for the ephemerides. In this way the measurements of
 Gliese 22 before 1995 were put into a consistent system of pixel
 scale and detector orientation.}
\begin{tabular}{lllllllll}
Epoch & Instrument & Telescope & \multicolumn{2}{c}{Measured}
 & \multicolumn{2}{c}{Predicted} &
  \multicolumn{2}{c}{Residuals} \\ 
 & & & \multicolumn{2}{c}{(calibrated with Trapezium)} &
  \multicolumn{2}{c}{(Scardia \cite{Sca83})} \\
  & & & PA [deg] & d [arcsec] & PA [deg] & d [arcsec] & $\Delta$PA &
  $\Delta$d \\ \hline
1995.526 & ESO NTT & SHARP\,II & 275.8 $\pm$ 0.1 & 1.848 $\pm$ 0.005 &
  276.5 & 1.868 & -0.7 & -0.020 \\
1996.641 & ESO 3.6-m & ADONIS & 274.5 $\pm$ 0.1 & 1.870 $\pm$ 0.008 &
  275.8 & 1.861 & -1.3 & 0.009 \\
1997.8843 & CA 3.5-m & MAGIC & 274.8 $\pm$ 0.2 & 1.838 $\pm$ 0.012 &
  275.0 & 1.853 & -0.2 & -0.015 \\
1999.6681 & CA 3.5-m & OMEGA Cass & 273.1 $\pm$ 0.1 & 1.852 $\pm$ 0.004 &
  273.8 & 1.842 & -0.7 & 0.010 \\ \hline
 & & & & & & mean residuals & -0.7 $\pm$ 0.2 & -0.004 $\pm$ 0.008 \\
\end{tabular}
\end{table*}

\section{Results}
\label{results}

\begin{table*}
\centering
\caption{\label{four_orbits}The orbits for Gliese~22~AC}
\begin{tabular}{lllll}
Element & Hershey (\cite{Her73}) & Heintz (\cite{Hei93}) &
  S\"oderhjelm (\cite{Sod99}) & This work \\ \hline
P (yr) & 15.95 $\pm$ 0.22 & 16.0 & 15.4 & 16.12 $\pm$ 0.2 \\
T      & 1956.0 $\pm$ 2.8 & 1988.9 & 1989.0 & 2000.47 $\pm$ 0.20 \\
$e$    & 0.05 $\pm$ 0.07  & 0.0    & 0.0    & 0.18 $\pm$ 0.03 \\
$a [``]$ & 0.51  & 0.525  & 0.450  & 0.529 $\pm$ 0.005 \\
$i$ [deg] & 45.0  &  42.0  & 27.0 &  46 $\pm$ 1 \\
$\Omega$  & 167.0 & 171.0  & 24.0 & 179.7 $\pm$ 1 \\
$\omega$  & 160.0 & 0.0 & 0.0 & 93.0 $\pm$ 5 \\ 
\hline
\end{tabular}
\end{table*}

\begin{table}
\centering
\caption{\label{ephem} Ephemerides for Gliese~22~AC}
\begin{tabular}{lll}
Epoch & $\theta$ [deg] & $\rho [``]$ \\ \hline
2003.0 & 351.9 & 0.485 \\
2004.0 & 8.5 & 0.521 \\
2005.0 & 24.2 & 0.517 \\
2006.0 & 41.1 & 0.489 \\
2007.0 & 60.3 & 0.455 \\
2008.0 & 82.1 & 0.435 \\
2009.0 & 104.7 & 0.441 \\
2010.0 & 125.5 & 0.469 \\
2011.0 & 143.4 & 0.506 \\
2012.0 & 159.2 & 0.531 \\
2013.0 & 174.3 & 0.527 \\
2014.0 & 190.8 & 0.483 \\
2015.0 & 212.7 & 0.400 \\
\end{tabular}
\end{table}

\begin{table}
\centering
\caption{\label{residuals} (O-C) residuals for our visual orbit for
 Gliese~22~AC and the most recent astrometric orbit for this system
 presented by S\"oderhjelm (\cite{Sod99}).
 $\theta$ is given in degrees and $\rho$ in arcseconds.}
\begin{tabular}{lllll}
Epoch & \multicolumn{2}{c}{S\"oderhjelm (\cite{Sod99})} &
        \multicolumn{2}{c}{This work} \\
        & $\theta$ & $\rho$ & $\theta$ & $\rho$ \\ \hline
1989.778  &  +2.4 & +0.006 & +3.5 & -0.041 \\
1989.939  &  +0.1 & +0.010 & +1.8 & -0.033 \\
1990.693  &  -3.3 & -0.026 & +0.6 & -0.055 \\
1991.717  & -13.2 & +0.032 & -6.8 & +0.007 \\
1993.745  & -14.7 & +0.056 & -0.3 &  0.000 \\
1994.074  & -16.2 & +0.065 & +0.1 & +0.001 \\
1994.704  & -19.5 & +0.085 & +0.6 & +0.010 \\
1994.947  & -21.1 & +0.102 & +0.7 & +0.024 \\
1995.767  & -26.9 & +0.082 &  0.0 & -0.003 \\
1996.739  & -32.7 & +0.083 &  0.0 & +0.003 \\
1997.876  & -38.7 & +0.020 & -0.6 & -0.023 \\
2001.107  &  -6.3 & -0.070 & +0.2 & +0.004 \\
2001.838  &  -1.6 & -0.012 & -0.8 & +0.009 \\
2002.797  &  -1.5 & -0.009 &  0.0 & -0.010 \\
\end{tabular}
\end{table}

\begin{figure*}
\centering
\includegraphics[width=8.5cm]{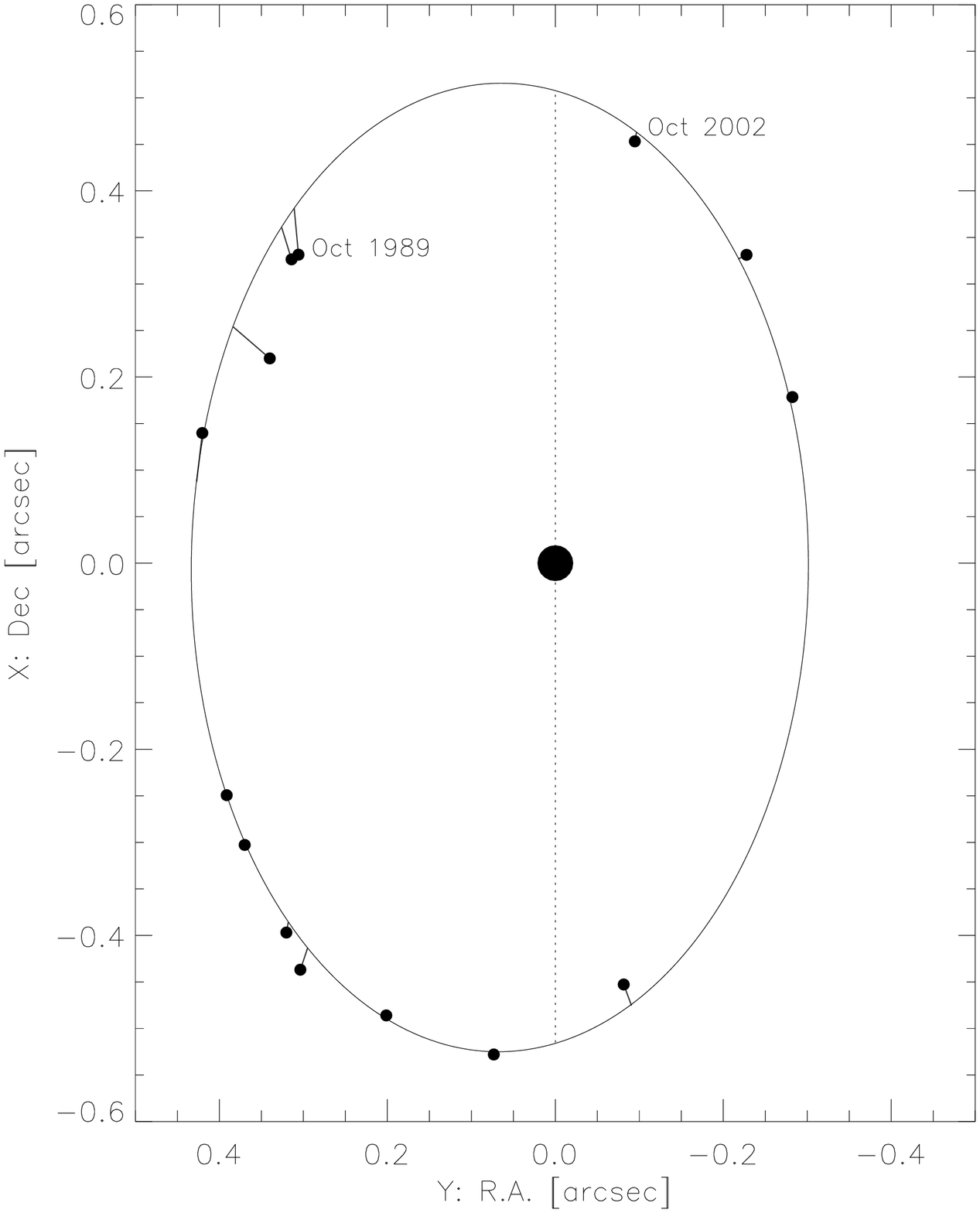}
\includegraphics[width=8.5cm]{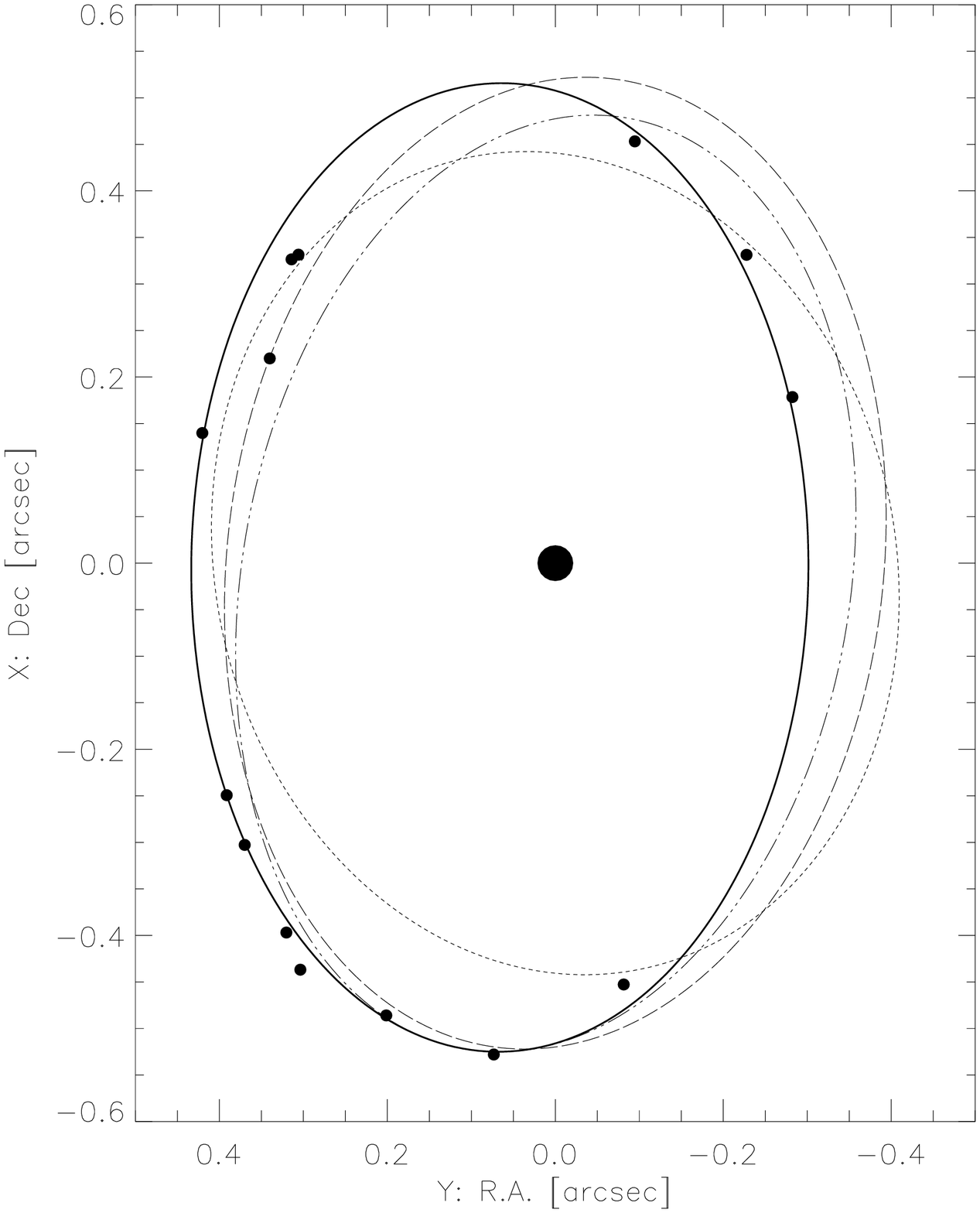}
\caption{\label{orbit_plot} Upper panel: Our visual orbit for Gliese 22 AC,
 together with the residuals of the individual data points. The line of
 nodes is also indicated. The motion of the companion is anticlockwise.
 Lower panel: Comparison of our data points with all four orbits from
 Table\,\ref{four_orbits}: Our orbit (thick solid line) and the
 astrometric orbits from Hershey (\cite{Her73}, dot-dot-dashed),
 Heintz (\cite{Hei93}, dashed) and S\"oderhjelm (\cite{Sod99}, dotted).}
\end{figure*}

To calculate the visual orbit indicated in Fig.\,\ref{orbit_plot},
we used the method proposed by Docobo (\cite{Doc85}).
A weight from 4 to 10 was assigned to the individual measurements
(see Table\,\ref{obs_AC_table}) to take into account seeing
conditions and the quality of different instruments.
The resulting orbital elements and their uncertainties are presented
in Table \ref{four_orbits} together with the results of the three previous
astrometric orbital solutions. Ephemerides for $\theta$ and $\rho$
until epoch 2015.0 are given in Table \ref{ephem}.
The residuals of our data points with respect to our visual orbit
and the most recent astrometric orbit by S\"oderhjelm \cite{Sod99}
(Table\,\ref{residuals}) indicate that our visual orbit represents a
significant improvement. This is also evident from Fig.\,\ref{orbit_plot} (right panel)
where our data points are plotted together with all four orbits.
The precision of the orbital elements is rather high, and we expect
only minor changes to this orbit. Thus, grade 2 (good orbit) can be
assigned to it according to the grading scheme described in the
Sixth Catalog of Orbits of Visual Binary Stars (Hartkopf \& Mason
\cite{Har03}).\\
The system mass derived from $a$ and $P$, using the {\it Hipparcos} parallax
(98.74 $\pm$ 3.37 milli-arcsec) is equal to $0.592\pm 0.065 M_{\sun}$.
This represents a rather high relative accuracy as the direct dynamical
mass sum determination concerned. While a good quality orbit (in
particular, $a$ and $P$ are determined with an accuracy of almost
1\%) is obtained, the accuracy of mass determination is still 11\%.
The latter is a good accuracy itself since the usual values for
post-Hipparcos epoch are in the order of $20-25$\% (see Martin et al.
\cite{Mar98}, S\"oderhjelm \cite{Sod99}) but it is worth noting that
it could be even better.\\
The principal reason in this (and many other) cases is the low accuracy
of the {\it Hipparcos} parallax (more than 3\%) whose contribution to
the mass error is 88\% while semimajor axis and period contribution
are almost insignificant (5\% and 7\% respectively). Gliese~22~AC
is thus a good example of the overall mass accuracy deterioration
due basically to the low relative accuracy of the parallax.
Therefore, one must state that the sensibly higher accuracy
next generation post-Hipparcos parallaxes are needed to drastically
improve the relative accuracy of direct mass determination. Should
semimajor axis, period and parallax be each determined with 1\%
accuracy, a mass determination accuracy of 5\% can be achieved
which is especially important for the lower end of luminosities in
the HR diagram.

\section{Discussion}
\label{disc}
We can now compare our dynamical mass with predictions from the present
mass-luminosity relations on the lower main sequence as well as from
theoretical evolutionary models for low-mass objcts.
The absolute (system) K band magnitude for Gliese~22~AC is
6.26~$\pm$~0.12\,mag (McCarthy et al.\,\cite{Mcc91})
The mean flux ratio from all measurements in
Table\,\ref{obs_AC_table} is $F_C/F_A = 0.156\pm  0.005$.
This results in absolute magnitudes for the components:

\begin{eqnarray}
\label{k_mag}
K_A & = & 6.42\pm 0.12 \nonumber \\
K_C & = & 8.43\pm 0.13 
\end{eqnarray}

The K band mass-luminosity relation given by Henry \& McCarthy
(\cite{Hen93}, their Eq.\,2) translates this into component masses:

\begin{eqnarray}
M_A = 0.378^{+0.028}_{-0.025} M_{\sun} \nonumber \\
M_C = 0.136^{+0.007}_{-0.007} M_{\sun}
\end{eqnarray}

The sum of these masses is $0.514\pm 0.029\,M_{\sun}$. The dynamical
system mass of $0.592 \pm 0.065\,M_{\sun}$ is in line with this
result only within its $2\sigma$ error. McCarthy et al.\,(\cite{Mcc91})
did a similar calculation and found that the mass sum derived
from the infrared magnitude-mass relations of Henry \& McCarthy
(\cite{Hen90}) and the dynamical mass were remarkably consistent.
For the latter they used however the orbit calculation by
Hershey (\cite{Her73}) which is worse than our visual orbit
(see Fig.\,\ref{orbit_plot}) and results in a probably too low system
mass of $0.485\pm 0.071\,M_{\odot}$. Our result indicates
that the the K band mass-luminosity relations by Henry \& McCarthy
probably underestimate stellar masses for M~dwarfs. A similar
result can be obtained for the {\it optical} mass-luminosity
relation given by Henry et al. (\cite{Hen99}). Using resolved
photometry taken with the HST Fine Guidance Sensors and
updated parallax information (van Altena et al. \cite{Alt95}),
these authors derive a system mass of $0.489 \pm 0.041\,M_{\odot}$.
This is again distinctly lower than our dynamical value.\\
The components' K band magnitudes from Eq.\,\ref{k_mag} can also
be used to derive masses from the theoretical evolutionary models
for low-mass stars and substellar objects presented by Baraffe
et al.\,(\cite{Bar98}, \cite{Bar02}). For this purpose one has to
assume an age for Gliese~22. Since this system shows no signs of
chromospheric activity (H$\alpha$ is in absorption, see Herbst
\& Miller \cite{Her89}),
we adopt a lower age limit of $10^8\,\mathrm{yr}$. As can be seen
from Table\,\ref{age_table}, the system mass derived from the theoretical
models is $\le 0.54\,M_{\sun}$ for all ages above this value.
This is again less than the dynamical system mass of $0.592\,M_{\sun}$,
but comparable to this empirical result within the uncertainties.
The restriction of this discussion to ages $\ge 10^8\,\mathrm{yr}$
causes no bias since lower ages would yield much lower (and thus
unrealistic) mass estimates.\\
Gliese~22~AC is a variable star that shows flare events with amplitudes
of 0.6~mag at optical wavelenghts (Pettersen \cite{Pet75}). One may
ask how this property influences the previous discussion. Even in
the unlikely case that the components' magnitudes from Table\,\ref{k_mag}
were affected by flares, our conclusions would not be altered.
Temporary higher luminosities would lead to higher mass estimates,
and thus the masses inferred from the mass-luminosity relation and
theoretical evolutionary models would be even lower than our values
if derived from observations in the quiescent state of a flare star.

\begin{table}
\centering
\caption{\label{age_table} Masses for Gliese~22~A and C derived
 from their K magnitudes (Eq.\,\ref{k_mag}) using the theoretical
 evolutionary models from Baraffe et al. (\cite{Bar98}, \cite{Bar02}).}
\begin{tabular}{llll}
Age [yr] & $M_A [M_{\sun}]$ & $M_C [M_{\sun}]$ & $M_{Sys} [M_{\sun}]$ \\ \hline
$10^8$     & 0.36 & 0.095 & 0.455 \\
$10^{8.5}$ & 0.40 & 0.14 & 0.54 \\
$10^9$     & 0.39 & 0.14 & 0.53 \\
$10^{10}$  & 0.39 & 0.14 & 0.53 \\ \hline
\end{tabular}
\end{table}  

The lower mass component Gl~22~C is apparently a very low-mass star with
a mass close to the hydrogen burning mass limit at $M\approx 0.075\,M_{\odot}$.
Therefore, it would be very interesting to derive its {\it individual}
dynamical mass from an absolute orbit and compare it to the
predictions of theoretical models. It will probably indeed be
possible to disentangle the relative orbit for Gliese~22~AC into
two absolute orbits for both components using the distant companion
Gliese~22~B as astrometric reference. However, to obtain most
reliable absolute orbits, a complete coverage of an orbital
revolution of the AC~pair is strongly desirable, and this is
thus beyond the scope of this paper.

\section{Summary}
With the first calculation of a purely visual orbit from interferometric
data we have largely improved on the accuracy of the orbital elements
for the low-mass binary Gliese~22~AC and derived a dynamical system mass
of $0.592\pm 0.065\,M_{\sun}$. The uncertainty of 11\% is mostly due
to the error of the {\it Hipparcos} parallax, while future ($\rho$, $\theta$)
measurements will probably result in only minor changes of the orbital
parameters. Based on the K band magnitudes of the components we have
estimated their masses also from the mass-luminosity relation given by
Henry \& McCarthy (\cite{Hen93}) and from the theoretical evolutionary
models for low-mass objects by Baraffe et al. (\cite{Bar98}, \cite{Bar02}).
In both cases the obtained mass sum is lower than the dynamical system mass.
The component Gliese~22~C is apparently a very low-mass star with a
mass around $0.1\,M_{\sun}$ and is thus located at the very end of the
lower main sequence.

\begin{acknowledgements}
We would like to thank Rainer K\"ohler for providing his software package
``Binary/Speckle'' for the reduction of 2D speckle-interferometric data.
J.W. acknowledges support from the Deutsches Zentrum f\"ur Luft- und
Raumfahrt under grant number 50 OR 0009. Visiting Observations on
Calar Alto were made possible by the Deutsche Forschungsgemeinschaft
under grant numbers Wo 834/1-1, Wo 834/2-1 and Wo 834/4-1.
This paper was supported by the grants AYA 2001-3073 of Spanish
Ministerio de Ciencia y Tecnologia and PGIDIT02 PXIC24301PN of
Xunta de Galicia. 
\end{acknowledgements}

\end{document}